\documentclass[12pt,a4paper]{article}
\usepackage{jheppub}

\usepackage{mathrsfs}

\usepackage[utf8]{inputenc}

\numberwithin{equation}{section}

\newcommand{\cB}{\mathcal{B}}

\newcommand{\cJ}{\mathcal{J}}

\newcommand{\cN}{\mathcal{N}}

\newcommand{\cR}{\mathcal{R}}
\newcommand{\cS}{\mathcal{S}}
\newcommand{\cT}{\mathcal{T}}

\newcommand{\cW}{\mathcal{W}}

\newcommand{\cZ}{\mathcal{Z}}

\renewcommand{\t}{\widetilde }
\renewcommand{\d}{\partial }
\renewcommand{\b}{\bar}

\newcommand{\D}{{\mathscr D}}
\newcommand{\J}{{\mathscr J}}

\newcommand{\LL}{{\mathscr{L}}}

\newif\ifpublic
\publicfalse
\ifpublic
\else
\RequirePackage{color}
\newcommand{\remark}[2][.]{{\color[rgb]{0,0,1}\renewcommand{\bfdefault}{b}\if.#1\else\textbf{#1:} \fi#2}}

\fi

\title{Defect multiplets of $\cN=1$ supersymmetry in 4d}

\author[a]{N.~Drukker}
\author[a]{I.~Shamir}
\author[a,b]{C.~Vergu}
\affiliation[a]{Department of Mathematics, King's College London \\
The Strand, WC2R 2LS, London, UK}
\affiliation[b]{Kavli Institute for Theoretical Physics\\
University of California\\
Santa Barbara, CA 93106, USA}

\abstract{%
Any 4d theory possessing $\cN=1$ supersymmetry admits a so called $\cS$-multiplet,
containing the conserved energy-momentum tensor and supercurrent. When a defect
is introduced into such a theory, the $\cS$-multiplet receives contributions localised on
the defect, which indicate the breaking of some translation symmetry and consequently
also some supersymmetries. We call this the defect multiplet. We classify such terms corresponding to half-BPS defects
which can be either three-dimensional, preserving 3d $\cN=1$, or
two-dimensional, preserving $\cN=(0,2)$. The new terms localised on the defect furnish
multiplets of the reduced symmetry and give rise to the displacement operator.
}

\begin{document}

\maketitle




\section{Introduction}

In this paper we study the energy-momentum multiplet of supersymmetric theories with defects. For such objects, localised on a submanifold $\Sigma$, the most robust feature is the pattern of space-time symmetry breaking, particularly the breaking of translations normal to the defect. This leads to a violation of the conservation of the energy-momentum tensor in the form
\begin{align} \label{bosonic_displacement}
\d^{\mu} T_{\nu\mu} = \delta(\Sigma) D_{\nu}
\end{align}
where $D_{\nu}$ is called the \emph{displacement operator} and has non-vanishing components orthogonal to the defect. Some applications of the displacement operator are as follows. The two-point
function of the displacement operator of Wilson lines determine the radiation of
an accelerated particle via the Bremsstrahlung function
\cite{Polyakov:2000ti, Semenoff:2004qr,Correa:2012at,Bianchi:2017ozk,Giombi:2017cqn}. The displacement operator can be used to determine the dependence of
entanglement entropy on the geometry of the entangling surface
\cite{Bianchi:2015liz,Dong:2016wcf, Balakrishnan:2016ttg, Bianchi:2016xvf, Chu:2016tps}.
It has also been used in the study of defect conformal field theories \cite{Mcavity:1993ue,Carlo,Herzog:2017kkj,Herzog:2017xha,Fukuda:2017cup,Armas:2017pvj} and to find constraints on defect field theories flows
\cite{Jensen:2015swa,Billo:2016cpy,Gadde:2016fbj}.

We focus on defects in 4d theories with $\cN=1$ supersymmetry which preserve half of the supersymmetry (defects in $\cN=1$ supersymmetric field theories have been recently studied in \cite{Maruyoshi:2016caf,Yagi:2017hmj,Closset:2017bse}; see also \cite{Gaiotto:2013sma} for a discussion the displacement operator in $\cN=2$ theories). For defects which are defined by space-like normal vectors there are two cases:
\begin{itemize}
\item
3d defects preserving an algebra isomorphic to $\cN=1$ in 3d. This is generated by linear combinations of supercharges of opposite chirality.
\item
2d defects preserving an algebra isomorphic to $\cN=(0,2)$ in 2d generated by choosing one component of the 4d supercharge $Q_\alpha$ and its complex conjugate, which we label $Q_+$ and $\bar Q_+$.
\end{itemize}

The energy multiplet of a 4d theory with $\cN=1$ supersymmetry resides in a $16+16$ multiplet known as the $\cS$-multiplet \cite{Komargodski:2010rb}.%
\footnote{Special theories can accommodate sub-multiplets, \emph{e.g.}\ a superconformal theory admits the $8+8$ superconformal multiplet \cite{Ferrara:1974pz,Gates:1983nr}.}
It is defined by the equation
\begin{align} \label{S_mult_intro}
\b D^{\dot\alpha} \cS_{\alpha\dot\alpha} = 2(\chi_\alpha - D_\alpha X),
\end{align}
where $\cS_{\alpha\dot\alpha}$ is a real vector superfield, $X$ and $\chi_\alpha$ are chiral superfields and
$D^\alpha \chi_\alpha = \b D_{\dot\alpha} \b \chi^{\dot\alpha}$. The superspace expansion of the $\cS$-multiplet is schematically
\begin{align} \label{S_mult_comp}
\cS_{\mu}
=
 j_\mu- i \theta^\alpha S_{\alpha\mu} + i \b \theta_{\dot\alpha} \b S^{\dot\alpha}{}_\mu + 2 \theta \sigma^\nu \b \theta T_{\nu\mu} + \ldots
\end{align}
where $T_{\nu\mu}$ is the conserved and symmetric energy-momentum tensor and $S_{\alpha\mu}$ is the conserved supercurrent.

The goal of this paper is to find additional terms that can arise on the right hand side of \eqref{S_mult_intro} due to the presence of defects. Such terms have delta-function support on the defect indicating the breaking of symmetry and leading to the displacement operator. In a previous paper \cite{Drukker:2017xrb} we focused on the case of 3d defects. In this paper we treat both 2d and 3d defects in the same unified framework (the differences in the preserved subalgebra nevertheless necessitate a separate treatment). To complement the previous paper, the examples we consider here are of 2d defects.

As motivation it is useful to think of $T_{\nu\mu}$ as 4 conserved global currents, and $\cS_\mu$ as 4 global current multiplets. The conservation of these multiplets is given by the 4 equations $\b D^2 \cS_\mu = 0$ up to total derivatives. What sort of terms can appear on the rhs of this equation? We are looking for delta-function terms that violate the conservation of the currents corresponding to translations normal to the defect. Such terms must be chiral.  This suggests the following structure
\begin{equation} \label{disp_mult_int}
\bar D^2 \mathcal{S}_\mu
=\begin{cases}
i\delta(\tilde y^n)n_\mu\Sigma,
\quad
&\text{3d defect,}\\
i\delta^{(2)}(y^{+-},y^{-+})\theta^- n^a_\mu \Sigma_{- a},
\quad
&\text{2d defect.}
\end{cases}
\end{equation}
Here $\Sigma$ and $\Sigma_{-a}$ are chiral operators, which are embeddings of 3 and 2d superfields in the  4d superspace, $n_\mu$ is the normal to the 3d defect and $n_\mu^a$ with $a=1,2$ are the  two normals of the 2d defect. The argument of the first delta function, $\tilde y^n = y^n - i\theta^2$, is a chiral combination which is simultaneously invariant under the 3d $\cN=1$ subalgebra. Similarly the combination $\delta^{(2)}(y^{+-},y^{-+})\theta^-$ is chiral and invariant under $Q_+$ and $\b Q_+$ (where $x^{+-}$ and $x^{-+}$ are directions normal to the defect). Since the terms of the right hand side are not total derivatives they violate the conservation and lead to the existence of a displacement operator.
In Section~\ref{Smult_defect_terms} we show which terms on the rhs of \eqref{S_mult_intro} produce these contributions to $\b D^2 \cS_\mu$.

As a simple illustration of equation \eqref{disp_mult_int} consider the following 3d example.
Consider a 4d chiral superfield $\Phi$ with an $x^n$ dependent coupling $\tau$ in
the superpotential. $\tau$ may be a $\delta$-function, but in fact the $\delta$-functions
in \eqref{disp_mult_int} can be any function representing the profile of a `smeared' defect.
To preserve two supercharges the coupling must take the form
$\tau(\tilde y^n) W(\Phi)$. For a canonical K\"{a}hler potential we have
$\cS_{\alpha\dot\alpha} = \b D_{\dot\alpha} \b\Phi D_\alpha \Phi$ which
leads, by using the equation of motion, to
\begin{align} \label{}
\b D^2 \cS_{\mu} = - 8i \tau(\tilde y^n) \d_\mu W.
\end{align}
Here the $\tilde y^n$ dependence of $\tau$ obstructs the rhs from being a
total derivative, thus preventing the existence of a conserved current in the
$n^\mu$ direction, encapsulating the main point of this paper.

The rest of this paper is organised as follows. In Section~\ref{superspace_emb} we describe in more detail the superspace embeddings and the invariant coordinates. Section~\ref{Smult_defect_terms} is the main part of the paper, in which we classify new defect terms that can be added to the $\cS$-multiplet, and derive from them the displacement operator. In Section~\ref{2d_example} we review the $\cN=(0,2)$ energy-momentum multiplet and show how it is embedded in the $\cS$-multiplet to account for the 2d degrees of freedom. In addition, we study two examples. The first consists of 2d Fermi multiplets which are coupled to 4d chiral multiplets. The second example consists of coupling a 4d gauge multiplet to charged 2d matter multiplets.


\section{Superspace embedding and invariant coordinates}
\label{superspace_emb}

As reviewed in the introduction new defect terms on the rhs of the $\cS$-multiplet equation must pass
several consistency conditions to respect the preserved superalgebra and chirality properties of the
equation. In this section we introduce the formalism that enables writing such terms. We introduce the
preserved 3d and 2d subalgebras and find appropriate coordinates to describe the tangent and
normal (super)spaces. The main outcome are delta functions localised on the defect which preserve
the subalgebras and have prescribed chirality properties.
These are the building blocks for the new defect terms in the $\cS$-multiplet equation, which are
constructed in Section~\ref{Smult_defect_terms}.

\subsection{Three dimensional defects}
\label{sec:3d-not}

A planar 3d defect is defined by a normal vector $n^\mu$. We use the the coordinate $x^n$ normal to the
defect and $x^i$ along it. Without loss of generality, the defect is then placed at $x^n=0$.
Such an object can preserve at most two supercharges given by the linear combinations
\(\hat Q_A = \tfrac 1 {\sqrt{2}}(\lambda_A^\alpha Q_\alpha + \bar{\lambda}_A^{\dot{\alpha}} \bar{Q}_{\dot{\alpha}})\)
for \(A = 1, 2\), where $(\lambda_A^\alpha)^* = \b \lambda_A^{\dot\alpha}$. The two-component $\hat Q_A$
is a real $SL(2,\mathbb{R})$ spinor of the 3d Lorentz group on the defect. In 4d we use
the conventions of Wess and Bagger \cite{Wess:1992cp} and treat 3d spinors as undotted 4d spinors.
Given 4d Weyl spinors $\psi_\alpha$ and $\b\psi_{\dot\alpha}$ it is convenient to define
\begin{align} \label{}
\psi_A = \lambda_A^\alpha \psi_\alpha,
\qquad
\b\psi_A = \b\lambda_A^{\dot\alpha} \b\psi_{\dot\alpha}.
\end{align}
We introduce $\lambda_\alpha^A$ and $\b \lambda_{\dot\alpha}^A$ with opposite index positioning, normalised by $i\lambda_A^\alpha \lambda^B_\alpha = -i\b\lambda^{\dot\alpha}_A \b\lambda^B_{\dot\alpha}=\delta_A^B$. This gives the inverse relations $\psi_\alpha = i \lambda_\alpha^A \psi_A$, $\b\psi_{\dot\alpha} = -i \b\lambda_{\dot\alpha}^A \b\psi_A$. Upper index spinors are given by
\begin{align} \label{}
\psi^A = \epsilon^{AB} \psi_B = - \lambda^A_\alpha \psi^\alpha,
\qquad
\b\psi^A = \epsilon^{AB} \b\psi_B = - \b\lambda^A_{\dot\alpha} \b \psi^{\dot\alpha}.
\end{align}
Finally, spinor bilinears are related by $\psi^A \chi_A = i \psi^\alpha \chi_\alpha$ and $\b \psi^A \b \chi_A = i \b\psi_{\dot\alpha} \b\chi^{\dot\alpha}$. Bilinears of real 3d spinors are imaginary.

In general, a four-vector \(v_\mu\) can be decomposed under the broken Lorentz subgroup as a scalar and a three-vector.  Explicitly, this reads
\begin{equation}
\label{decompose}
\lambda_A^\alpha \bar{\lambda}_B^{\dot{\alpha}} v_{\alpha \dot{\alpha}}
= \left( \lambda_{(A}^\alpha \bar{\lambda}_{B)}^{\dot{\alpha}}
+ \frac{1}{2} \epsilon_{A B} \lambda^{\alpha C} \b\lambda_{C}^{\dot\alpha} \right) v_{\alpha\dot\alpha}
= \Gamma_{A B}^i v_i + i \epsilon_{A B} v^n ,
\end{equation}
where $\Gamma_{A B}^i = \Gamma_{B A}^i$ are the 3d Dirac matrices, and $i\lambda_\alpha^{A} \b \lambda_{\dot\alpha A} = n_{\alpha\dot\alpha}$ is the normal vector. We sometimes use $v_{AB} \equiv \Gamma^i_{AB} v_i$.
It can now be easily checked that the anti-commutator is given by \(\{\hat Q_A, \hat Q_B\} = 2i \d_{AB} \). Therefore, the supercharges \(\hat Q_A\) are part of the three-dimensional super-Poincar\'e subalgebra preserved by the defect.%

It is natural to construct superspace coordinates which are adapted to the action of the 3d subalgebra. In general the 4d supersymmetry transformations are
\begin{gather}
\delta \theta^\alpha = \zeta^\alpha, \qquad
\delta \bar{\theta}^{\dot\alpha} = \bar{\zeta}^{\dot\alpha}, \qquad
\delta x^\mu = i \theta \sigma^\mu \bar{\zeta} - i \zeta \sigma^\mu \bar{\theta},
\end{gather}
and the subalgebra is identified via the relation $\zeta^A = \b\zeta^A$. Let us then define Grassmannian coordinates $\Theta^A$ transverse to the defect and $\t\Theta^A$ normal to it by
\begin{equation}
\Theta^A = \frac 1 {\sqrt{2}} (\theta^A + \bar{\theta}^{A}),
\qquad
\t\Theta^A = \frac i {\sqrt{2}} (\theta^A - \bar{\theta}^{A}).
\end{equation}
Then, under the transformations generated by \(-i \zeta^A \hat{Q}_A\) we have the variations
\begin{equation}
\delta \Theta^A = \zeta^A, \qquad
\delta \t\Theta^A = 0, \qquad
\delta x^n
= i \zeta^A \t\Theta_A.
\end{equation}
Note that $\t\Theta^A$ is an invariant coordinate. In the same way, the modification of $x^n$
to \(\tilde{x}^n = x^n - i \t\Theta \Theta\) is also invariant. We then
have a natural basis for the 4d superspace $(x^i, \Theta^A; \tilde{x}^n, \t\Theta^A)$ which is
adapted to the 3d subalgebra. Clearly $(x^i,\Theta^A)$ can be identified with the 3d superspace.
Furthermore, since \(\t\Theta^A\) is invariant, we can add any terms depending only on
\(\t\Theta^A\) to the definition of \(\tilde{x}^n\) while preserving the invariance property.
In fact, one can make a chiral combination
\begin{equation}
\tilde{y}^n = x^n - i \t\Theta \Theta - \t\Theta^2
= y^n - i \theta^2.
\end{equation}
Using this variable we can define a chiral invariant delta function $\delta(\tilde y^n)$,
which is used below to write the new terms on right hand side of the $\cS$-multiplet equation.

We also introduce covariant derivatives corresponding to our new coordinates by
\begin{equation}
\Delta_A = \frac 1 {\sqrt{2}} (D_A + \bar{D}_A),
\qquad
\t{\Delta}_A = -\frac {i}{\sqrt{2}} (D_A - \bar{D}_{A}).
\end{equation}
which satisfy the relations
\begin{equation}
\begin{aligned} \label{}
&\{ \Delta_A, \Delta_B \} = -2 i \d_{AB},
&\qquad
&\{ \Delta_A, \t \Delta_B \} =  2 i \epsilon_{AB} \d_n,
\\
&\Delta_A \Delta_B = - i \d_{AB} + \frac{1}{2} \epsilon_{AB} \Delta^2,
&\qquad
&\Delta_A \Delta^2 = - \Delta^2 \Delta_A = 2i \d_{AB} \Delta^B. \label{Delta_identities}
\end{aligned}
\end{equation}
We have the following useful formulas
\begin{equation}
\begin{aligned}
&\Delta_A \Theta^B = \t\Delta_A \t\Theta^B = i\delta_A^B,
&\qquad
&\Delta_A \t\Theta^B = \t{\Delta}_A \Theta^B = 0,
 \\
&\Delta_A \tilde x^n = 2 \t\Theta_A,
&\qquad
&\t{\Delta}_A \tilde x^n = 0.
\end{aligned}
\end{equation}

It is useful to study how superfield representations decompose under the preserved subgroup.
Working with the coordinates $(x^i, \Theta^A; \tilde{x}^n, \t\Theta^A)$ it is clear that the
component fields in the $\t\Theta^A$ expansion transform independently. A more useful
way of saying this is that the different sub-multiplets are obtained by the formula\footnote{An expression \(X\vert_{\widetilde{\Theta} = 0}\) stands for the superfield \(X\) evaluated at \(\widetilde{\Theta} = 0\), while the rest of the odd coordinates are kept arbitrary.  We also adopt the convention that \(X\vert\) is the bottom component of the superfield \(X\), i.e.\ we set all the odd arguments of \(X\) to zero.}
$\t \Delta_A^{\ell}(\ldots) |_{\t\Theta=0}$ where $\ell$ can be $0,1$ or $2$.
In the simplest case, of a chiral superfield \(X\), the first sub-multiplet is $X|_{\t\Theta=0}$
and it is then clear from $\sqrt{2}\bar{D}_{A} X = (\Delta_A - i \tilde{\Delta}_A) X = 0$
that all the other sub-multiplets can be determined from it.

Conversely, 3d multiplets can be embedded in the 4d superspace in various ways. For example,
consider a real 3d multiplet $C(x^i, \Theta^A)$. It has a chiral embedding given by
$C(y^i, \sqrt 2 \theta^A) = -\tfrac i 2 \b D^2 ( \t\Theta^2 C)$ (see \cite{Drukker:2017xrb} for other embeddings).

\subsection{Two dimensional defects} \label{sec:2d_defect}

For the codimension two case, we denote one of the supercharges by \(Q_+ = \lambda_+^\alpha Q_\alpha\)
and its hermitian conjugate by \(\bar{Q}_+ = \bar{\lambda}_+^{\dot{\alpha}} \bar{Q}_{\dot{\alpha}}\). This choice breaks the 4d Lorentz symmetry to $SO(1,1)\times SO(2)$, where $Q_+$ and $\b Q_+$ both have charge 1 with respect to $SO(1,1)$. There are unique supercharges $Q_-$ and $\b Q_-$ with charge $-1$, describing the broken supersymmetries, which are specified by $\lambda_-^\alpha$ and $\b \lambda_-^{\dot\alpha}$. The spinor variables $\lambda_+$ and $\lambda_-$ suggest a natural basis for the coordinates
\begin{align} \label{}
x_{\pm\pm} = x_{\alpha\dot\alpha} \lambda^\alpha_{\pm} \b\lambda^{\dot\alpha}_{\pm},
\qquad
x_{+-} = x_{\alpha\dot\alpha} \lambda^\alpha_{+} \b\lambda^{\dot\alpha}_{-},
\qquad
x_{-+} = x_{\alpha\dot\alpha} \lambda^\alpha_{-} \b\lambda^{\dot\alpha}_{+}.
\end{align}
Here $x_{\pm\pm}$ are the directions along the defect and $x_{\pm\mp}$ are the two normal directions.%
\footnote{For the defect in the $(x^0,x^3)$ plane, this agrees with the usual bi-spinor
notations $x_{\alpha\dot\alpha}$ so $\lambda$ and $\bar\lambda$ identify $1$ with $-$ and $2$ with $+$.}
The supercharges $Q_+$ and $\b Q_+$ satisfy
\begin{align} \label{}
\{ Q_+, \b Q_+ \} = 2i \d_{++},
\end{align}
which is the 2d $\cN=(0,2)$ algebra. We fix the phase of $\lambda_\pm$ so that $\lambda^\alpha_+ \lambda_{\alpha-} = 1$ and define superspace coordinates according to $\theta_\pm = \lambda_\pm^\alpha \theta_\alpha$ and $\b\theta_\pm=\b \lambda_\pm^{\dot\alpha} \b\theta_{\dot\alpha}$. Upper spinor indices are defined by $\theta^\pm= \mp \theta_\mp$ and $\b\theta^\pm = \mp \b\theta_{\mp}$. The covariant derivatives are given by%
\footnote{Note that in our conventions $\d_{\pm\pm} x^{\pm\pm} = \d_{\pm\mp} x^{\mp\pm} = -2$, which gives the non-standard formula $f(x^{--} - 2i \theta^- \b \theta^-) = f(x^{--}) + i \theta^- \bar{\theta}^- \d_{--} f(x^{--})$.}
\begin{gather}
D_\pm = \frac \partial {\partial \theta^\pm} + i \bar{\theta}^\pm \partial_{\pm \pm} + i \bar{\theta}^{\mp} \partial_{\pm\mp},
\qquad
\bar{D}_\pm = -\frac \partial {\partial \bar{\theta}^\pm} - i \theta^\pm \partial_{\pm\pm} - i \theta^\mp \partial_{\mp\pm},
\end{gather}
satisfying the algebra
\begin{gather}
\{D_{\pm}, \bar{D}_{\pm}\} = -2 i \partial_{\pm\pm}, \qquad
\{D_{\pm}, \bar{D}_{\mp}\} = -2 i \partial_{\pm\mp}.
\end{gather}

The complex conjugate 2d supersymmetry parameters $\zeta^+$ and $\bar\zeta^+$
can be identified with half of those in 4d, such that the supersymmetry transformations are
\begin{gather}
\delta \theta^+ = \zeta^+, \qquad
\delta \bar\theta^+ = \bar\zeta^+, \qquad
\delta x^{++}=-2i(\theta^+\bar\zeta^+ - \zeta^+\bar\theta^+).
\end{gather}
To parametrise the normal directions one can use $(x^{+-}, x^{-+}, \theta^-,\bar\theta^-)$. But while $\theta^-$ and $\b \theta^-$ are invariant under the 2d supersymmetry, the even coordinates are not. It is thus useful to define the invariant bosonic coordinates
\begin{align} \label{2d_inv_coord}
\tilde x^{+-} = x^{+-} - 2i \b\theta^+ \theta^-, \qquad
\tilde x^{-+} = x^{-+} + 2i \b\theta^- \theta^+.
\end{align}
These invariant coordinates are annihilated by $D_-$ and $\b D_-$.
It is easy to check that $\tilde x^{-+}$ is chiral (in fact $y^{-+} = \tilde x^{-+}$), but its complex
conjugate $\tilde x^{+-}$ is anti-chiral.
In fact, it is not possible to change $\tilde x^{+-}$, by adding Grassmann bilinears, to a chiral quantity while preserving its invariance.

The terms on the right hand side of the defect multiplet equation should be annihilated by
$\bar D^2$. Those associated to the defect can be written as $\delta^{(2)}(\tilde x^{+-},\tilde x^{-+})K$,
with $K$ a superfield. The non-chiral delta function is annihilated by $\b D_-$ (and therefore also $\b D^2$) and its $\bar D_+$ derivative is proportional to $\theta^-$,
so $K$ has to satisfy $\b D^2 K=0$ and $\b D_- K|_{\theta^-=0}=0$ (note that the second condition
does not imply the first because of the projection $\theta^-=0$). One simple solution to these
conditions is $K=\theta^-\cZ_-$, where now we only need to impose $\bar D^2\cZ_-=0$. In fact
if we adjoin $\theta^-$ to the delta function, the combination
$\delta^{(2)}(\tilde x^{+-},\tilde x^{-+})\theta^-$ is chiral and invariant. The chirality follows from $\theta^- f(\tilde x^{+-}) = \theta^- f(y^{+-})$ for any function $f$. We indeed find such terms
in the examples we study, but this is not the only solution, and we also find $K$ with nonzero
bottom component. As we discuss in~\ref{sec:gauge-defect}
the conditions on $K$ are precisely those defining the embedding of a 2d $\cN=(0,2)$ multiplet of conserved
current in the corresponding 4d multiplet.


\section{New defect terms in the \texorpdfstring{$\cS$}{S}-multiplet}
\label{Smult_defect_terms}

We now come to the main section of this paper, where we analyse the defect-supported
terms that can appear on the rhs of the $\cS$-multiplet equation \eqref{S_mult_intro}.
The presence of the defect breaks translation symmetry and half the supersymmetries. The new delta-function supported terms on the rhs of \eqref{S_mult_intro} violate some of the conservation laws, but should leave the currents associated with the preserved subalgebra conserved.

To that end we consider the possible singular terms on the rhs and decompose them
into multiplets of the preserved subalgebra, and consequently require that those
containing the components of the supercurrent and energy-momentum
tensor of the defect theory be preserved.

\subsection{3d defects}

Let us start with the 3d case already studied in \cite{Drukker:2017xrb}.
We present it here for completeness and to highlight the analogy with 2d defects, discussed in the next section.

In the presence of a 3d defect the $\cS$-multiplet equation \eqref{S_mult_intro} is modified
to
\begin{equation} \label{Z3d}
\b D^{\dot\alpha} \cS_{\alpha\dot\alpha} = 2(\chi_\alpha - D_\alpha X)
+ i \delta(\tilde y^n)\lambda_\alpha^A\cZ_A.
\end{equation}
The last term has explicit dependence on the direction normal to the defect, breaking
the full 4d symmetry. The preserved symmetry allows for explicit dependence only on the invariant coordinates $\tilde y^n$ and $\t\Theta$, while the dependence on $x^i$ and $\Theta$ enters only via the fields.
We may expand in $\t\Theta$ to obtain the form \begin{equation}
\label{Z-expand}
\cZ_A=\Pi_A+\t\Theta_A\Pi - i \t\Theta^B\Pi_{AB}\,.
\end{equation}
We might naively expect the $\Pi$ superfields to have no $\t\Theta$ dependence, due to their 3d origin. In fact these superfields represent embeddings of 3d superfields in the 4d superspace and thus acquire a $\t\Theta$ dependence (see below). We therefore need some conditions on the $\Pi$'s to make the expansion unambiguous.
It follows from \eqref{Z3d} that $\cZ$ must satisfy a chirality condition. Since $\tilde y^n$ is
chiral, $\cZ$ should be annihilated by $\bar D^2$, and we require the $\Pi$'s to satisfy this
condition independently. This is the reason why we do not consider a $\t\Theta^2$ term in
\eqref{Z-expand}, and also requires us to allow for $\t\Theta$ dependence in the $\Pi$'s.

The considerations so far do not fix the $\Pi$'s uniquely, but there is in fact a unique choice for this decomposition with chiral $\Pi$'s. To see that, consider $\bar\lambda^{\dot\alpha}_B\bar D_{\dot\alpha}\cZ_A$. This is a chiral field which can be decomposed as in \eqref{decompose} to a scalar and a vector. This is how we choose $\Pi$ and $\Pi_{AB}$, which automatically makes $\Pi_A$ chiral as well. As discussed at the end of Section~\ref{sec:3d-not}, the chirality of these fields means that they do depend on $\tilde\Theta$, but this dependence is completely determined by their 3d coordinates $(x^i, \Theta_A)$.

Conservation of the 3d $\cN=1$ algebra imposes conditions on the $\Pi$ fields. To
understand them, it is useful to begin by considering the embedding of a decoupled 3d
energy-momentum multiplet into the 4d $\cS$-multiplet. Any 3d theory with $\cN=1$
supersymmetry admits an energy-momentum multiplet $\cJ_{Ai}$, in the 3d superspace $(x^i, \Theta_A)$, satisfying
\cite{Drukker:2017xrb}
\begin{align} \label{3d_J_mult}
\Delta^A \cJ_{Ai} = - 2i \d_i X,
\qquad
\Gamma^i{}_A{}^B \cJ_{B i} =  \Delta_A(X - Y),
\end{align}
where $X$ and $Y$ are real multiplets (the conformal version of this multiplet
with $X=Y=0$ was also considered in
\cite{Kuzenko:2011xg, Kuzenko:2010rp, Buchbinder:2015qsa}).
Using \eqref{Delta_identities} it is immediate to derive
$\d^i \cJ_{Ai} = (\Gamma^i \Delta)_A \d_i X $, which means that the bottom
component of $\cJ_{Ai}$ contains a conserved Majorana supercurrent.
In fact $\cJ_{Ai} = S_{Ai} - 2 (\Gamma^j \Theta)_A T_{ji} + \ldots$
where $S_{Ai}$ and $T_{ij}$ are the supercurrent and the energy-momentum
tensor respectively. Let us define the embedding of $\cJ_{Ai}$ in the $\cS$-multiplet by
\begin{align} \label{S3_emb1}
\cS^{(3)}_i
&=
i \delta(\tilde x^n) \t \Theta^A \cJ_{Ai}
=
\frac 1 {\sqrt{2}} (\theta^\alpha \lambda_\alpha^A - \bar{\theta}^{\dot{\alpha}} \bar{\lambda}_{\dot{\alpha}}^A) S_{A i} + 2\delta(x^n) \theta \sigma^j \b\theta T_{ji}  + \ldots
\end{align}
and $\cS^{(3)}_n = 0$.
Notice that this term combines with \eqref{S_mult_comp} to give the sum of energy-momentum contributions from the bulk and defect $T_{\nu\mu} + \delta(x^n) \delta^j_\nu \delta^i_\mu T_{ji}$. Plugging $\cS^{(3)}_\mu$ back in the $\cS$-multiplet equation gives us a first glimpse of what the defect terms should look like. A somewhat tedious computation gives
\begin{align} \label{S3_embedding}
\b D^{\dot\alpha} \cS^{(3)}_{\alpha\dot\alpha}
=
i \delta(\tilde y^n ) \lambda_\alpha^A \left( \b D^2 D_A \left( - \frac{i}{4}\t\Theta^2 (X-Y)\right)
- \sqrt2 i \t\Theta^B \d_{AB} X \right).
\end{align}
Let us make two comments about this expression. First, in the second term $X$ stands for the chiral embedding of the 3d multiplet (see Section~\ref{superspace_emb}) and comparison with \eqref{Z-expand} gives the identification $\Pi_{AB} = \sqrt2 \d_{AB}X$. Similarly, the first term is chiral and corresponds to $\Pi_A$. It projects to $\Pi_A |_{\t\Theta=0} = \Delta_A(X-Y)/\sqrt2$. Second, despite the appearance of a delta function, these terms do not lead to a displacement operator, as expected since a decoupled theory is insensitive to the location of the defect. A simple way of seeing this is to note that $\b D^2 \cS^{(3)}_{\mu} = - 2 i \delta(\tilde y^n) \delta_{\mu}^i \d_i X$ is a total derivative (compare with equation \eqref{disp_mult_int}).%
\footnote{In a purely 3d case we drop the delta function and take $\cS^{(3)}_i = i \t \Theta^A \cJ_{Ai}$. The analog of \eqref{S3_embedding} can be arranged to the form
\begin{equation*}
\b D^{\dot\alpha} \cS^{(3)}_{\alpha\dot\alpha}
=
\frac{1}{4} \b D^2 D_\alpha \left( i\t \Theta^2 Y\right)
- \frac{1}{4} D_\alpha \b D^2 \left( i \t \Theta^2 X \right).
\end{equation*}
This provides a more transparent relation with the 4d $\cS$-multiplet where the first term takes the form of $\chi_\alpha= - \frac{1}{4} \b D^2 D_\alpha V$ for some real $V$ and the second of $D_{\alpha} X$ from \eqref{S_mult_intro}.
}

Looking at \eqref{S3_emb1}, we identify the sub-multiplet of $\cS_\mu$ containing the currents which survive the symmetry breaking. It is given by
\begin{align} \label{3d_sub_S_mult}
-\t \Delta_A \cS_\mu |_{\t\Theta=0}
=
\frac{1}{\sqrt2}(\lambda_A^\alpha S_{\alpha \mu}  + \bar{\lambda}_{A}^{\dot{\alpha}} \bar{S}_{\dot{\alpha}\mu})
- 2 (\Gamma^j \Theta)_A T_{j \mu} + \dots
\end{align}
Notice that the current index is 4d. The new defect term in \eqref{Z3d} must be such
that $\t \Delta_A \d^\mu \cS_\mu |_{\t\Theta=0}$ remains a total derivative. The derivative of this reads, in terms of the expansion in \eqref{Z-expand}
\begin{align} \label{3d_conserv_condition}
-\tilde{\Delta}_A \partial^\mu \mathcal{S}_\mu |_{\t\Theta=0}
&=
\delta(x^n) \left. \left( \frac{i}{2\sqrt2} \Delta^B \Delta_A (\Pi_B + \bar{\Pi}_B) - \frac{1}{2\sqrt2} \Delta^B (\Pi_{AB} + \bar{\Pi}_{AB}) \right)\right|_{\t\Theta=0}
\nonumber \\
&\quad
+ \frac{i}{\sqrt2} \d_n \left. \left( \delta(x^n) (\Pi_A - \bar{\Pi}_A) \right) \right|_{\t\Theta=0}.
\end{align}
Conservation of the 3d sub-multiplet requires that the rhs is a total derivative. To analyse the conditions for this it is interesting to compare this with the following 3d computation. Consider again the 3d multiplet $\cJ_{Ai}$ in \eqref{3d_J_mult}, but this time let us be ignorant about the terms on the rhs and try instead
\begin{align} \label{}
\Delta^A \cJ_{Ai} = - i \Pi'_i,
\qquad
\Gamma^i{}_A{}^B \cJ_{B i}  = \Pi'_A,
\end{align}
where $\Pi'_i$ is a real vector and $\Pi'_A$ a real spinor. The condition for the conservation is now
\begin{align} \label{3dcon}
\d^i \cJ_{A i } = \frac{i}{2} \Delta^B \Delta_A \Pi'_B - \frac{1}{2} \Delta^B \Pi'_{AB}.
\end{align}
The analogy with \eqref{3d_conserv_condition}, except for the 4d normal derivative, should be obvious. The rhs must be a total derivative which suggests \eqref{3d_J_mult}, namely $\Pi'_A = \Delta_A (\ldots)$ and $\Pi'_{AB} = \d_{AB} (\ldots)$.\footnote{It follows from the identity \(\Delta^A \Delta_B \Delta_A = 0\) that the first term in \eqref{3dcon} vanishes for this choice.}
To be sure, this is not the most general solution of this equation. However it is sufficiently general in the same sense that the $\cS$-multiplet is -- any 3d theory with $\cN=1$ admits such a multiplet up to improvements. In the same way, a sufficiently general solution to \eqref{3d_conserv_condition} is the following: the real parts of $\Pi_A|_{\t\Theta=0}$ and $\Pi_{AB}|_{\t\Theta=0}$ should (at least locally) take the form $\Delta_A(\ldots)$ and $\d_{AB} (\ldots)$ respectively. The imaginary parts of $\Pi_A$ and $\Pi_{AB}$ are unconstrained and do not have a purely 3d origin. The imaginary parts together with $\Pi$ appear only as a result of interactions with 4d and are related to the reduced symmetry.

What is left is to extract the displacement operator.
The sub-multiplet containing the currents whose conservation is
violated due to the presence of the defect is given by
\begin{equation} \label{non_conserv_multiplet}
\mathcal{S}_\mu\vert_{\t\Theta = 0}
= -\frac{1}{\sqrt2} \Theta^A (\lambda_A^\alpha S_{\alpha \mu}
- \bar{\lambda}_{A}^{\dot{\alpha}} \bar{S}_{\dot{\alpha} \mu})
+ i \Theta^2 T_{n \mu} + \dots
\end{equation}
In terms of the expansion in \eqref{Z-expand} we find
\begin{equation}\label{eq:partial_s}
\partial^\mu \mathcal{S}_\mu\vert_{\t\Theta = 0}
=
-\frac{1}{2\sqrt2} \delta(x^n) \Delta^A (\Pi_A - \bar{\Pi}_A)|_{\t\Theta = 0}
- \frac{1}{2\sqrt2} \delta(x^n) (\Pi + \b \Pi)|_{\t\Theta = 0}.
\end{equation}
Since $T_{n \mu}$ appears in the $\Theta^2$ component of $\cS_\mu$ only $\Pi$
contributes to the displacement. We find
\begin{align} \label{}
\d^\mu T_{n \mu} = \frac{i}{4\sqrt2} \delta(x^n) \Delta^2 (\Pi + \bar\Pi)|
\end{align}
up to total derivatives. Finally, we can compare this result with the intuition
presented in the introduction. Clearly we have
$\b D^2 \cS_\mu = -\sqrt2 i \delta(\tilde y^n) n_\mu \Pi$ as anticipated.

\subsection{2d defects}

The 2d case works along the same lines as the 3d case, where now the $\cS$-multiplet
equation takes the form%
\footnote{We henceforth abbreviate the delta functions $\delta^{(2)}(x^{+-}, x^{-+})$, $\delta^{(2)}(\tilde{x}^{+-}, \tilde{x}^{-+})$ and $\delta^{(2)}(y^{+-},y^{-+})$ to $\delta^{(2)}(x)$, $\delta^{(2)}(\tilde x)$ and $\delta^{(2)}(\tilde y)$, respectively.}
\begin{equation} \label{Z2d}
\b D^{\dot\alpha} \cS_{\alpha\dot\alpha} = 2(\chi_\alpha - D_\alpha X)
+ \delta^{(2)}(\tilde x)(\lambda_\alpha^+\cZ_+ + \lambda_\alpha^-\cZ_-).
\end{equation}
Acting with $\b D^2$ on this equation we find \(0 = \bar{D}^2 (\delta^{(2)}(\tilde{x}) \mathcal{Z}_\pm)\).  As explained in Section~\ref{sec:2d_defect}, using \(\bar{D}_- \delta^{(2)}(\tilde{x}) = 0\) and \(\bar{D}_+ \delta^{(2)}(\tilde{x}) \propto \theta^-\), we conclude that \(\bar{D}^2 \mathcal{Z}_\pm = 0\) and \(\bar{D}_- \mathcal{Z}_\pm\vert_{\theta^- = 0} = 0\).

Let us now solve these constraints on \(\mathcal{Z}_\pm\).  The second constraint can be solved by \(\bar{D}_- \mathcal{Z}_\pm = \theta^- \Pi_{--\pm}\).  The left-hand side is annihilated by \(\bar{D}_-\) and by \(\bar{D}_+\) (due to the first constraint) so it follows that \(\Pi_{--\pm}\) is a chiral superfield.  Since in the cases of interest for us \(\Pi_{--\pm}\) only appears multiplied by \(\theta^-\) we can further impose \(D_- \Pi_{--\pm}\vert_{\bar{\theta}^- = 0} = 0\).

Let us now solve the equation \(\bar{D}_- \mathcal{Z}_\pm = \theta^- \Pi_{--\pm}\).  The general solution can be written as a linear superposition of a particular solution of the non-homogeneous equation and the solution of the homogeneous equation.  A solution of the non-homogeneous equation is \(\theta^- \bar{\theta}^- \Pi_{--\pm}\). The general solution to \(\bar{D}_- \mathcal{Z}_\pm = 0\) can be written as \( \Pi_\pm + i \theta^- \Pi_{-\pm}\), where \(\bar{D}_- \Pi_\pm = \bar{D}_- \Pi_{-\pm} = 0\) and we impose \(D_- \Pi_\pm\vert_{\bar{\theta}^- = 0} = 0\), \(D_- \Pi_{-\pm}\vert_{\bar{\theta}^- = 0} = 0\).

Hence, we have shown that
\begin{equation}
\label{new_2d_terms}
  \cZ_\pm = \Pi_\pm+ i\theta^- \Pi_{-\pm} + \theta^- \bar \theta^- \Pi_{--\pm},
\end{equation}
where
\begin{equation}
  \bar D_-\Pi_\star =D_- \Pi_\star\vert_{\bar{\theta}^- = 0} = 0,
\end{equation}
where \(\star\) can be any of the \(\pm\), \({-}{\pm}\) and \({-}{-}{\pm}\). The superfield $\Pi_{--\pm}$ is additionally annihilated by $\bar D_+$, so it is chiral.

As in the three-dimensional case, we proceed to examine how the $\cS$-multiplet splits into sub-multiplets
of the reduced symmetry.
We defer the discussion of the 2d energy-momentum multiplet and its embedding in the $\cS$-multiplet to Section~\ref{2d_example}, where we also present several examples illustrating the general result.

Upon introducing the defect, the translations along the
directions $x^{\pm\pm}$ remain unbroken. This means that the currents
$T_{\pm\pm \mu}$ are conserved, while \(T_{+- \mu}\) and \(T_{-+ \mu}\) are not.
Likewise the supercurrents $S_{+\mu}$ and $\b S_{+\mu}$ which correspond to
$Q_+$ and $\b Q_+$ are conserved, while $S_{-\mu}$ and $\b S_{-\mu}$ are not.
Starting from the Grassmann expansion of the \(\mathcal{S}\) multiplet
in~\eqref{S_mult_comp}, we find the sub-multiplets
\begin{align} \label{}
&\cS_\mu |_{\theta^- = 0}
=
j_\mu - i \theta^+ S_{+\mu} + i \b \theta^+ \b S_{+ \mu} +2 \theta^+ \b \theta^+ T_{++\mu} + \ldots,
\label{bot_S_comp_3} \\
& [D_-, \b D_-] \cS_\mu |_{\theta^- = 0}
=
4 T_{--\mu} + \ldots,
\label{top_S_comp_3}\\
&D_- \cS_\mu |_{\theta^- = 0}
=
-i S_{-\mu} +  \b\theta^+ (2T_{-+ \mu} + i \d_{-+} j_\mu )+ \ldots. \label{middle_S_comp_3}
\end{align}
The first two multiplets include the currents that should remain unbroken, though the first
includes also $j_\mu$, the $R$-current, which may be broken. The $\Pi$
terms \eqref{new_2d_terms} on the rhs of the $\cS$-multiplet equation must be
consistent with this requirement.

Consider first the conservation of the currents in
\eqref{bot_S_comp_3}. It is verified by computing%
\footnote{A useful first step to obtain this is to take $\b D^{\dot{\alpha}} \cS_{\alpha\dot\alpha} = \cB_\alpha$ for some $\mathcal{B}_\alpha$, then $4i \d^\mu \cS_\mu = D^\alpha \cB_\alpha - \b D_{\dot\alpha} \b \cB^{\dot\alpha}$. From this it is also possible to get the useful expressions
\begin{align*} \label{}
D_\alpha \d^\mu \cS_\mu
&=
\frac{i}{8} \left( D^2 \cB_\alpha - 2\b D_{\dot\alpha} D_{\alpha} \b \cB^{\dot\alpha} \right) + \text{total derivative,}
\\
[D_\alpha, \b D_{\dot\alpha}] \d^\mu \cS_\mu
&=
- \frac{i}{4} \left( D^2 \b D_{\dot\alpha} \mathcal{B}_\alpha + \b D^2 D_\alpha \bar{\mathcal{B}}_{\dot\alpha} \right) + \text{total derivative,}
\end{align*}
}
\begin{align} \label{}
\d_\mu \cS^\mu |_{\theta^- = 0}
=
 -\frac 1 4 \delta^{(2)}(x)\left. \left( \Pi_{-+} + \b \Pi_{-+} + i D_+ \Pi_- + i \b D_+ \b \Pi_- \right)\right|_{\theta^- = 0}.
\end{align}
All but the bottom component of this expression should be total derivatives.
In the 2d $\cN=(0,2)$ superspace (which is where this equation effectively lives after
setting $\theta^-=0$) this can be achieved by requiring that the term in the brackets
be written as $D_+ \cW_- - \b D_+ \b\cW_-$ where $\cW_-$ is an $\cN=(0,2)$ chiral
superfield. The $R$-current is conserved if $\cW_- = \b D_+ \cR_{--}$ for a real
$\cR_{--}$. Note that the imaginary parts of \(\Pi_{-+}\) and $i D_+ \Pi_-$ are unconstrained. Similarly, to check the conservation of the currents in \eqref{top_S_comp_3}, we consider
\begin{align} \label{cons_condition_2d}
[D_-, \b D_-] \d_\mu \cS^\mu |_{\theta^- = 0}
&=
 \delta^{(2)}(x) \left.\left( -\frac{i}{2} D_+ \Pi_{---} + \text{c.c.} \right)\right|_{\theta^-=0}
+ \text{total derivative}.
\end{align}
Since the bottom component of \eqref{top_S_comp_3} is the energy-momentum tensor $T_{--\mu}$ the rhs of \eqref{cons_condition_2d} must be a total derivative.
Lastly, to obtain the displacement operator we compute
\begin{align} \label{eq:displ-op-generic}
D_- \partial^\mu\cS_\mu |_{\theta^- = 0}
&=
-\frac{1}{4} \delta^{(2)} (x) \left( D_+ \Pi_{--} - i \b \Pi_{--+} \right)|_{\theta^-=0} + \text{total derivative}.
\end{align}
It is instructive to compare this with the other approach to finding the displacement operator,
in the introduction \eqref{disp_mult_int}. We find
\begin{equation}
\begin{aligned} \label{}
\b D^2 \cS_{+-}
&=
2 \delta^{(2)}(y) \theta^- \Pi_{--+},
\\
\b D^2 \cS_{-+}
&=
-2i \delta^{(2)}(y) \theta^- \b D_+ \Pi_{--} + 2 \b D_+ \left( \delta^{(2)}(\tilde x) \Pi_- \right).
\end{aligned}
\end{equation}
The second term on the second line leads to a total derivative and does not contribute to
the displacement. To compare to~\eqref{eq:displ-op-generic}, note that
\eqref{middle_S_comp_3} leads to $\d^\mu T_{-+ \mu} = - \frac{1}{2} \b D_+ D_- \d^\mu \cS_\mu |$. On the other hand we have the relation
\begin{equation}
\begin{aligned} \label{}
\d^\mu T_{\mu -+}
&=
- \frac{i}{32} [D^2, \b D^2] \cS_{-+} |
\\
&=
\frac{1}{8} \delta^{(2)}(x) \left( \b D_+ D_+ \Pi_{--} - i \b D_+ \b \Pi_{--+} \right)\big| + \text{total derivative},
\end{aligned}
\end{equation}
which gives the same result up to total derivatives. In general the energy-momentum
tensor does not have to be symmetric since some of the Lorentz symmetries are
broken, but up to total derivatives, these expressions lead to the same answer.


\section{Examples of 2d defects}
\label{2d_example}

In this section we present a few examples of 2d defects which are defined by
coupling a bulk 4d theory to an $\cN=(0,2)$ theory living on a 2d subsurface. Specifically,
we construct the energy-momentum multiplets of these defect field theories, which include a bulk contribution as well as a contribution localised on the defect, and compute the displacement multiplet. This illustrates the general discussion of Section~\ref{Smult_defect_terms}.

The first example is that of a theory of 2d \(\mathcal{N} = (0,2)\) Fermi multiplets, interacting with bulk chiral multiplets. The second example is that of a 2d chiral multiplet interacting with a 4d gauge multiplet.

To motivate these examples and to keep our presentation self-contained we start by reviewing some background material regarding $\cN=(0,2)$ in 2d. We present the energy-momentum multiplet of such theories \cite{Dumitrescu:2011iu} and construct it explicitly for several example.

\subsection{The energy-momentum of 2d \texorpdfstring{$\cN=(0,2)$}{N=02} theories}
\label{sec:2d-em-tensor}

The 2d superspace has coordinates $(x^{\pm\pm}, \theta^+, \b\theta^+)$. The conventions here are obtained from our 4d conventions in Section~\ref{superspace_emb} by dimensional reduction. The energy-momentum multiplet is defined by \cite{Dumitrescu:2011iu}
\begin{align} \label{}
\d_{--} \cJ_{++} = D_+ \cW_- - \b D_+ \b \cW_-,
\qquad
\b D_+ \cT_{----} = - \d_{--} \cW_-,
\end{align}
where $\cJ_{++}$ and $\cT_{----}$ are real multiplets and $\cW_-$ is chiral, \emph{i.e.\@} $\b D_+ \cW_-=0$. It is straightforward to show that the energy-momentum tensor defined by
\begin{equation}
\begin{gathered} \label{}
T_{----} = \cT_{----}|,
\qquad
T_{++++} = \frac{1}{4} [D_+, \b D_+] \cJ_{++}|,
\\
T_{--++} = T_{++--} = \frac{i}{2}(D_+\cW_- + \b D_+ \b \cW_-)|
\end{gathered}
\end{equation}
is conserved. Similarly $S_{+++} =iD_+ J_{++}| $ and $S_{+--} = 2 \b\cW_- |$ are components of a conserved supercurrent.%
\footnote{If $\cW_-= -\frac{i}{2} \b D_+ \cJ_{--}$ for a real (and well defined) multiplet $\cJ_{--}$, then the theory has a conserved $R$-current. If $\cW_-=0$ then the theory is superconformal.}

\paragraph{Chiral superfields}

The simplest 2d model is that of chiral superfield $\b D_+ \varphi = 0$, which can be obtained from a 4d chiral $\Phi$ by the projection $\varphi = \Phi|_{\theta^-=0}$. In components it is given by
\begin{align} \label{}
\varphi = \phi + \sqrt2 \theta^+ \psi_+ + i\theta^+ \b \theta^+ \d_{++} \phi.
\end{align}
Let us consider the following Lagrangian
\begin{align} \label{2d_free_chiral}
\LL_{\text{chiral}}
=
-\frac{i}{4} \int d\theta^+ d\b \theta^+\left( \b\varphi \d_{--} \varphi - \d_{--} \b\varphi \varphi \right),
\end{align}
which leads to the equation of motion $D_+ \d_{--} \varphi=0$.
The energy-momentum multiplet of this model is given by
\begin{align} \label{}
\cJ_{++}
=
\b D_+ \b\varphi D_+ \varphi,
\qquad
 \cT_{----} = 2 \d_{--} \b\varphi \d_{--} \varphi,
 \qquad
 \cW_- = 0.
\end{align}

\paragraph{Gauge fields}
\label{par:gauge-fields}

Anticipating the application to 4d defect theories, we present the gauged version of
the above model where the gauge multiplet comes from dimensional reduction.
Recall, that the 4d gauge multiplet is contained in a real superfield \(V\), which in
the Wess-Zumino gauge reads
\begin{equation}
  V = -\theta \sigma^\mu \bar{\theta} v_\mu + i \theta^2 \bar{\theta} \bar{\lambda} - i \bar{\theta}^2 \theta \lambda + \frac 1 2 \theta^2 \bar{\theta}^2 D.
\end{equation}
We need the sub-multiplets which contain only $v_{++}$ and $v_{--}$. They are given by $G = V|_{\theta^-=0}$ and $G_{--} = - \frac{1}{2} [D_-, \b D_-] V|_{\theta^-=0}$. Specifically, (in the Wess-Zumino gauge)
\begin{align} \label{}
G = -\theta^+ \b\theta^+ v_{++},
\qquad
G_{--} = v_{--} + 2i \theta^+ \b\lambda_- + 2i \b\theta^+ \lambda_- -2 \theta^+ \b\theta^+ D.
\end{align}
The relevant sub-multiplet of the 4d field strength \(W_\alpha = -\frac 1 4 \bar{D}^2 D_\alpha V\) can be defined by $\Upsilon_- =W_-|_{\theta^-=0}=\frac{1}{2} \b D_+ (G_{--} - i \d_{--} G)$. This yields
\begin{equation}
\begin{aligned}
\Upsilon_-
= -i \lambda_- - \left(D + i F_{-+}\right) \theta^+ + \theta^+ \b\theta^+ \d_{++} \lambda_-, 
\end{aligned}
\end{equation}
where \(F_{-+} = \frac{1}{2}(\partial_{++} v_{--} - \partial_{--} v_{++})\).

To couple the gauge field to the chiral multiplet let us define gauge covariant derivatives by
(see \cite{Witten:1993yc})
\begin{equation}
\begin{aligned} \label{}
&\D_{++} = \d_{++} + i v_{++},
\qquad&
&\D_{--} = \d_{--} + i G_{--},
\\
&\D_+ = \frac{\d}{\d \theta^+} + i \b\theta^+ \D_{++},
\qquad&
&\b\D_+ = -\frac{\d}{\d \b\theta^+} - i \theta^+ \D_{++},
\end{aligned}
\end{equation}
which satisfy the algebra $\{ \D_+, \b \D_+ \} = - 2 i \D_{++}$ and
$[ \b \D_+, \D_{--} ] = 2i \Upsilon_-$. A chiral superfield is now defined by the constraint $\b \D_+ \varphi = 0$.  The gauge covariant derivatives replace the usual derivatives in the Lagrangian \eqref{2d_free_chiral}. This leads to the equation of motion
$\D_+ \D_{--} \varphi = 0$ (the order of the derivatives is now important).

The Lagrangian for gauge multiplet is
\begin{equation}
  \mathscr{L}_{\text{gauge}} = - \frac 1 {2 g^2} \int d \theta^+ d \bar{\theta}^+ \Upsilon_- \bar{\Upsilon}_-,
\end{equation}
which leads to the equations of motion
\begin{align} \label{}
D_+ \Upsilon_- - \b D_+ \b \Upsilon_- = 2 g^2 \J,
\qquad
i \d_{--} \left( D_+ \Upsilon_- + \b D_+ \b \Upsilon_- \right) = - 2g^2 \mathscr{J}_{--}.
\end{align}
Here $\mathscr{J} = \b\varphi \varphi$ and $\mathscr{J}_{--} = i(\varphi \D_{--} \b\varphi-\b\varphi \D_{--} \varphi)$ constitute the current multiplet whose conservation is given by $\b D_+ (\mathscr{J}_{--} + i \d_{--} \mathscr{J}) = 0$. (Note that due to the constraint on the chiral superfield a variation of $G$ has to be accompanied by $\delta \varphi = \delta G \varphi$). The energy-momentum multiplet of this theory is given by
\begin{equation}
\begin{gathered} \label{2d_gauged_chiral}
\cJ_{++} = \b\D_+ \b\varphi \D_+ \varphi,
\qquad
\cW_- = \frac{i}{g^2 } \Upsilon_- \b D_+ \b \Upsilon_-,
\\
\cT_{----} = 2 \D_{--} \b\varphi \D_{--}\varphi - \frac{i}{g^2} \left(\Upsilon_- \b \d_{--} \b \Upsilon_- - \d_{--} \Upsilon_- \b \Upsilon_-\right).
\end{gathered}
\end{equation}

\paragraph{Fermi multiplets}

A Fermi multiplet $\Lambda_-$ is defined by the relation $\b D_+ \Lambda_- =  E$ where $E$ is a holomorphic function of chiral multiplets
$\varphi^I$. Its components are
\begin{align} \label{}
\Lambda_- = \psi_- - \sqrt2 \theta^+ F +i \theta^+ \b \theta^+ \d_{++} \psi_- -  \b\theta^+ E.
\end{align}
From a 4d point of view, any chiral superfield $\Phi$ decomposes into a 2d chiral
$\varphi = \Phi|_{\theta^-=0}$ and a Fermi multiplet
$\Lambda_- = \frac{1}{\sqrt2} D_- \Phi |_{\theta^-=0}$, in which case
$E = - \sqrt2 i \d_{-+} \varphi$.

For the system of Fermi multiplets interacting with chirals, one adds $\LL_\text{chiral}$ of \eqref{2d_free_chiral} to the Lagrangian
\begin{align}
  \mathscr{L}_\text{F} +   \mathscr{L}_J  = \frac 1 2 \int d \theta^+ d \bar{\theta}^+ \bar{\Lambda}_-^a \Lambda_{-,a}
 + \frac 1 {2} \int d \theta^+ \Lambda_{-,a} J^a + \text{c.c}.
\end{align}
Here \(\bar{D}_+ \Lambda_{-,a} = E_a\), with \(E_a\) and \(J^a\) holomorphic
functions of \(\varphi^I\) subject to the constraint \(E_a J^a = 0\). We find the equations of motion
\begin{equation}
\bar{D}_+ \partial_{--} \bar{\varphi}_I = -i (\bar{\Lambda}_-^a \partial_I E_a + \Lambda_{-,a} \partial_I J^a),
\qquad
\b D_+ \b \Lambda_-^a = J^a.
\end{equation}
It is easy to see that for a free Fermi multiplet (with $E=J=0$) the only contribution to the energy-momentum tensor is $T_{----} = - i(\b\psi_- \d_{--} \psi_- - \d_{--} \b \psi_- \psi_-)$.
This suggests the following energy-momentum multiplet
\begin{equation}
\begin{gathered}
  \mathcal{J}_{++} =  \bar{D}_+ \bar{\varphi}_I D_+ \varphi^I , \qquad
  \mathcal{W}_- = i \left(\bar{\Lambda}_-^a E_a + \Lambda_{-,a} J^a\right), \\
  \mathcal{T}_{----} = 2 \partial_{--} \b \varphi_I \partial_{--} \varphi^I  -  i \left(\bar{\Lambda}_-^a \partial_{--} \Lambda_{-,a} - (\partial_{--} \bar{\Lambda}_-^a) \Lambda_{-,a}\right).
\end{gathered}
\end{equation}

\subsection{Systems with 2d-4d interactions}
\label{sec:fermi-defect}

We now consider $\cN=1$ theories in 4d interacting with degrees of freedom localised on a 2d plane. If we assume that the theory on the defect admits $\cN=(0,2)$ supersymmetry then it is straightforward to write interactions which preserve this supersymmetry. This is achieved by decomposing the bulk fields into $\cN=(0,2)$ sub-multiplets as indicated above. The sub-multiplets can then be coupled in the standard way consistent with the $\cN=(0,2)$ symmetry on the defect.

As discussed in Section~\ref{Smult_defect_terms}, the combined system has an energy-momentum tensor which includes, in addition to the usual bulk terms, also contributions localised on the defect. Unless the defect degrees of freedom decouple, only the total energy-momentum is conserved and only in the directions parallel to the defect. In particular, in the supersymmetric case, it is clear that the 2d energy-momentum multiplet has to be embedded in the bulk $\cS$-multiplet. On the other hand, if the defect degrees of freedom decouple, their contribution to the energy-momentum multiplet has to satisfy the $\cS$-multiplet equation by itself. Therefore as preparation we study the embedding of the 2d multiplet in the $\cS$-multiplet. It is given by
\begin{equation}
\begin{aligned} \label{}
&\cS_{++} = \cJ_{++},
&\qquad
&\cS_{--} = 2 \theta^- \b \theta^- \cT_{----},
\\
&\chi_- = \frac{i}{2} \cW_- (y^{\pm\pm}, \theta^+)
&\qquad
&\chi_+ = -\frac{i}{2} \theta^- \b D_{+} \b \cW_-,
\\
&X = \frac{i}{2} \theta^- \cW_- (y^{\pm\pm}, \theta^+).
\end{aligned}
\end{equation}
Here $\cS_{++} = \lambda^\alpha_+ \b\lambda_+^{\dot\alpha} \cS_{\alpha\dot\alpha}$ etc. and $\cS_{\pm\mp}=0$.
Strictly speaking $\cW_-(y^{++}, x^{--}, \theta^+)$ is a 2d chiral, \emph{i.e.}, annihilated by $\b D_+$. It is here embedded via $\cW_-(x^{--}) \to \cW_-(y^{--})$ such that it is a full 4d chiral. It is not obvious that $\chi_+$ as defined above is chiral with respect to $\b D_-$. Nevertheless, since $\cW_-$ is a 2d field it follows that $\b D_-$ effectively acts as $- i \theta^- \d_{--}$, which vanishes due to the explicit factor of $\theta^-$ in $\chi_+$. This can be made more explicit by rewriting it as $\chi_+ = - \frac{i}{4}\b D^2 \left( \theta^- \b \theta^- \b \cW_-\right)$. Finally, since we can always locally solve $\cW_- = i \b D_+ \omega_{--}$ for some real multiplet $\omega_{--}$, it follows that $\chi_{\alpha} = -\frac{1}{4}\b D^2 D_\alpha(\theta^-\b\theta^- \omega_{--})$ and therefore $\chi_\alpha$ satisfies the Bianchi identity.

\paragraph{Chiral-Fermi defect}

Consider (free) 4d chiral superfields $\Phi^I$ and Fermi multiplets \(\Lambda_{-,a}\) which live on a 2d defect.

We denote the restriction of \(\Phi\) to two dimensions by \(\varphi\). Then let $\varphi$ and $\Lambda_-$ interact on the defect by introducing $E_a$ and $J^a$ following the prescription in the previous section. To obtain the 4d equation of motion of $\Phi^I$ we note the relation
\begin{align} \label{}
\int dy^{++} dx^{--} \int d\theta^+ \delta \varphi (\ldots)
=
-2 \int d^4y \int d^2\theta \delta^{(2)}(y) \theta^-\delta \Phi (\ldots)
\end{align}
where the ellipses correspond to some chiral field.
The equations of motion are then given by
\begin{equation}
\begin{gathered}
  \bar{D}_+ \Lambda_{-,a} = E_a(\varphi), \qquad
  \bar{D}_+ \bar{\Lambda}_-^a = J^a, \\
  \bar{D}^2 \bar{\Phi}_I = - 4 \delta^{(2)}(y) \theta^- \left(\bar{\Lambda}_-^a \partial_I E_a+ \Lambda_{-,a} \partial_I J^a \right).
\end{gathered}
\end{equation}
Note that the chiral delta function appears in a combination with $\theta^-$ which makes it also invariant under the preserved subalgebra, as it must be.

The contribution to $\cS$-multiplet from 4d and 2d can be written as follows
\begin{equation}
\begin{aligned} \label{}
\cS^{(4)}_{\alpha\dot\alpha}
&=
\b D_{\dot\alpha} \b \Phi_I D_\alpha \Phi^I,
\\
\cS^{(2)}_{--}
&=
-2 i \delta^{(2)}(\tilde x) \theta^- \b \theta^- \left( \b\Lambda_-{}^a \d_{--} \Lambda_{-a} - \d_{--} \b\Lambda_-{}^a \Lambda_{-a} \right),
\end{aligned}
\end{equation}
with $\cS^\text{tot} = \cS^{(4)} + \cS^{(2)}$. It is then straightforward to compute $\b D^{\dot\alpha} \cS^\text{tot}_{\alpha\dot\alpha}$, to obtain the various $\Pi$ terms and to show explicitly that the conditions for conservation are upheld. We instead just proceed directly to the displacement multiplet. As explained in the introduction, the quick way of obtaining it is by computing
\begin{align} \label{}
- \frac{1}{4}\b D^2 \cS^\text{tot}_{+-} = 4 i \delta^{(2)}(y) \theta^- \left( \b\Lambda^a_- \d_{+-} E_a + \Lambda_{-,a} \d_{+-} J^a\right),
\end{align}
and similarly for $\cS_{-+}$. Clearly, only $\cS^{(4)}$ enters this computation. Since the rhs is not a total derivative $T_{+-\mu}$ is not conserved.

\paragraph{Bulk gauge field with charged matter on the defect}
\label{sec:gauge-defect}

As a second example, take a gauge field arising from a bulk 4d theory and couple it to the 2d chiral $\varphi$.  The 2d gauge multiplet $(G,G_{--})$ is induced from the bulk 4d multiplet $V$ by $G=V|_{\theta^-=0}$ and $G_{--} = - \frac{1}{2} [D_-,\b D_-] V|_{\theta^-=0}$.
The equations of motion for the 2d chiral superfields have been derived above, in the paragraph on gauge fields~\ref{par:gauge-fields}.

To derive the bulk gauge field equation of motion we proceed as follows. In general, the 4d gauge multiplet satisfies the Bianchi identify $D^\alpha W_\alpha = \b D_{\dot\alpha} \b W^{\dot\alpha}$ and the equation of motion $D^\alpha W_\alpha + \b D_{\dot\alpha} \b W^{\dot\alpha} = 4J$ where $J$ is a conserved current, \emph{i.e.}, $\b D^2 J = D^2 J=0$, normalised so that $J= \b\Phi\Phi$ where $\Phi$ has unit charge. For the 2d chiral $\varphi$ the conserved current multiplet is comprised of two parts $(\mathscr{J},\mathscr{J}_{--})$ with the conservation equation being $\b D_+ (\J_{--} + i \d_{--} \J)=0$. The 2d multiplet can be embedded in 4d by defining
\begin{align} \label{}
J = \delta^{(2)}(\tilde x) \left( \J - \theta^- \b\theta^- \J_{--} \right),
\end{align}
which is easily checked to be conserved in the 4d superspace sense. Since $\J = \b\varphi \varphi$ where $\varphi$ has unit charge it is correctly normalised.

The 4d part of the defect multiplet for the gauge fields is as usual $\cS^{(4)}_{\alpha\dot\alpha} = 2W_\alpha \b W_{\dot\alpha}$. The 2d parts can be taken directly from \eqref{2d_gauged_chiral}
\begin{align} \label{}
\cS^{(2)}_{++} = \delta^{(2)}(\tilde x)\b \D_+ \b\varphi \D_+ \varphi,
\qquad
\cS^{(2)}_{--} = 4 \delta^{(2)}(\tilde x)\theta^- \b\theta^- \D_{--} \b\varphi \D_{--}\varphi.
\end{align}
From this we obtain
\begin{equation}
\begin{aligned} \label{}
\b D^{\dot\alpha} \cS^\text{tot}_{-\dot\alpha}
&=
4\delta^{(2)}(\tilde x) \left( \J  + i \theta^- \b\theta^- \d_{--}\J \right) W_-
\\
&=
4\delta^{(2)}(\tilde x) \left( \J \Upsilon_- + i \theta^- \b\theta^- \d_{--}(\J\Upsilon_-) \right)
+ 4\delta^{(2)}(y)\theta^- \J f_{--},
\end{aligned}
\end{equation}
where $W_- (y^{\pm\pm},\tilde x^{\pm\mp}) = \Upsilon_- + f_{--} \theta^-$. The first term here corresponds to $\Pi_-$ and indeed satisfies the chirality condition $\b D_- \Pi_-=0$. The second term corresponds to $\Pi_{--}$ which contributes to the displacement operator. The second component of the defect multiplet equation is
\begin{equation}
\begin{aligned} \label{}
\b D^{\dot\alpha} \cS^\text{tot}_{+ \dot\alpha}
&=
4\delta^{(2)}(\tilde x) \left( \J \Upsilon_+ + i \theta^- \b\theta^- \d_{--}(\J\Upsilon_+) \right)
 \\
&\quad +\delta^{(2)}(y)\theta^- i \left( D_+ \cW'_- - \b D_+ \b \cW'_- -2i \J ( f_{+-} + \b f_{+-} )\right)
\\
&\quad - 4\delta^{(2)}(y) \theta^- \b\theta^- (\J_{--} + i \d_{--} \J) \Upsilon_+,
\end{aligned}
\end{equation}
with $\cW'_- = -2i \J \Upsilon_-$. The term in the first line corresponds to $\Pi_+$, the second to $\Pi_{-+}$ whose real part (in the brackets) is of the form $D_+ \cW'_- - \b D_+ \b \cW'_-$ and imaginary part is unconstrained. The third line is the chiral $\Pi_{--+}$ which gives the second piece in the displacement operator.
It is straightforward to obtain $\b D^2 \cS_{\pm\mp}$ which gets contributions only from the 4d part of $\cS^{(4)}_{\alpha\dot\alpha}$. We find
\begin{equation}
\begin{aligned} \label{}
\b D^2 \cS_{+-} &= -8 \delta^{(2)}(y) \theta^- \Upsilon_+ (\J_{--} + i \d_{--} \J),
\\
\b D^2 \cS_{-+} &=  8 \b D_+  \left(\delta^{(2)} (\tilde x)\Upsilon_- \J(y^{--}) \right)
-8 \delta^{(2)}(y) \theta^- f_{--} \b D_+ \J.
\end{aligned}
\end{equation}
In the second line we used $\J(y^{--}) = \J + i \theta^- \b\theta^- \d_{--} \J$. The displacement operator includes the following bosonic contribution
\begin{align} \label{}
\d^\mu T_{\mu -+}
&=
\frac{1}{2} \delta^{(2)}(x)\left. \bigg((\J_{++} + i\d_{++} \J) F_{---+} + (\J_{--}- i\d_{--} \J) F_{++-+}  \bigg) \right|
\\
&\quad \text{+ fermions.}
\end{align}
where $\J_{\pm\pm}$ are the components of the 2d conserved current, \emph{i.e.}, $\J_{++} = -\frac{1}{2} [D_+ , \b D_+] \J |$, and $F_{\mu\nu} = \d_\mu v_\nu - \d_\nu v_\mu$.

\section*{Acknowledgments}

It is a pleasure to thank Matt Buican, Dario Martelli and Daisuke Yokoyama for illuminating discussions. This research was supported in part by the National Science Foundation under Grant No.~NSF PHY-1125915 and by the Science \& Technology Facilities Council via the consolidated grant numbers ST/J002798/1 and ST/L00326/1

\bibliographystyle{JHEP}
\bibliography{bibliography}

\end{document}